\documentstyle[graphicx,multicol,prl,aps]{revtex}

\newcommand {\qdc} {{q^{2}_c}}

\begin{document}


\title{Correlation Time Scales in the Sherrington-Kirkpatrick Model}

\author{Alain Billoire}

\address{
  CEA/Saclay,
  Service de Physique Th\'eorique,
  91191 Gif-sur-Yvette, France.
}

\author{Enzo Marinari}

\address{
  Dipartimento di Fisica, INFN and INFM, 
  Universit\`a di Roma {\em La Sapienza},\\
  P. A. Moro 2, 00185 Roma, Italy. }

\date{\today}                                                 

\maketitle

\begin{abstract}  
We investigate the dynamical behavior of the Sherrington-Kirkpatrick
mean field model of spin glasses by numerical simulation. All the time
scales $\tau_x$ we have measured behave like $\ln(\tau_x) \propto
N^{\epsilon}$, where $N$ is the number of spins and $\epsilon\simeq
\frac13$. This is true whether the autocorrelation function used to
define $\tau_x$ is sensitive to the full reversal of the system or not.
\end{abstract}


\begin{multicols}{2}
\narrowtext
\parskip=0cm

Today many features of the Sherrington-Kirkpatrick mean field model of
spin glasses \cite{BINYOU,MEPAVI,FISHER} have been clarified.
Probably most questions that still need investigations are related to
the very interesting dynamics of the model (see for example
\cite{AGING}). Here, following Mackenzie and Young \cite{MACYOU}, we
examine the equilibrium dynamics of the model. In this classic paper
the authors gave numerical evidences, from systems with up to 192
spins, for the existence of a spectrum of relaxation times which
diverge with the number of spins $N$ as $\ln(\tau) \propto
N^{\frac14}$, and of a second, longer ``ergodic'' time scale
$\tau_{eg}$ which is the time needed to turn over all the spins, with
$\overline{<\ln( \tau_{eg})>} \propto N^{\frac12}$.  For doing that
one looks both at processes that require a full reversal of all the
spins and at processes that on the contrary are not sensitive to this
phenomenon. In this letter, we establish that indeed all dynamical
scales have the same behavior, compatible with barrier heights growing
like $N^{\epsilon}$, where $\epsilon\simeq 0.3$ close to the
$N^{\frac13}$ behavior suggested in \cite{RODMOO,VERVIR} (see also the
numerical simulations in \cite{COLBOR}).


Let us start by giving some details about our simulation.  We study
systems with $N=64$, $128$, $256$, $512$, and $1024$ spins, with $\pm
1$ couplings.  We first thermalize the system using the {\em parallel
tempering} optimized Monte Carlo procedure \cite{PARTEM} with a set of
$38$ $T$ values in the range $0.4-1.325$ (i.e. $\Delta T = 0.025$). We
perform $400000$ iterations (one iteration consists of one Metropolis
sweep plus one tempering update cycle), and store the final well
equilibrated configurations. Next we start updating these equilibrium
configurations (more precisely the subset with $T$ = 0.4, 0.5 , \dots)
with a simple Metropolis dynamics, and perform $4 \cdot 10^6$
Metropolis sweeps. We have in all cases two replica and $512$
realizations of the disorder.

For each of these samples we compute the {\em flip
times}  $\tau_1$, $\tau_2$ and $\tau_3$. We define $\tau_1^{(J)}$ as
the time after which, on a given sample $J$, the time dependent
self-overlap

\begin{equation}
q(0,t)\equiv\frac1N \sum_i \sigma_i(0) \sigma_i(t)
\end{equation}
has become smaller than $+\Sigma$, with

\begin{equation}
  \Sigma \equiv \sqrt{\langle q^2 \rangle_{J}}\ ,
\end{equation}
where $\langle q^2 \rangle_{J}$, the usual square Parisi overlap, is
computed during the second half of the thermalization run for the
given sample. The time $t$ is measured in units of sweeps, with $t=0$
at the beginning of the Metropolis dynamics.  We define analogously
$\tau_2$ as the time it takes to $q(0,t)$ to decay from its initial
value of $1$ down to $0$, and $\tau_3$ as the time it takes to
$q(0,t)$ to decay down to $-\Sigma$.

We expect\footnote{Notice that while $\tau_2$ and $\tau_3$ are
unambiguous signatures of the transition to the reversed part of the
phase space, $\tau_1$ can be ambiguous, since depending on $T$ it can
still characterize a transition in the short time regime or already an
ergodic transition. The fact that the three $\tau_i$ turn out to be
compatible gives further support to the existence of a single time
scale exponent $\epsilon$.} $\tau_1$, $\tau_2$ and $\tau_3$ to obey
the same scaling law. In the following we will try to check if an
exponential scaling of the kind

\begin{equation}
  \tau_{1,2,3} \simeq A_{1,2,3} \exp\left(\alpha_{1,2,3} N^{\epsilon}\right)
  \label{E-TAUSCALE}
\end{equation}
gives a good fit to the data, and we will try to determine
$\epsilon$.

We base our analysis on empirical {\em medians} for $\ln(\tau)$, i.e. we
sort the 512 values of $\ln(\tau)$ as $\ln(\tau^{(0)}) \leq
\ln(\tau^{(1)}) \leq \dots \leq\ln(\tau^{(511)})$ (more precisely the
512 values of $\ln(\tau)$ averaged over the two replica) and define the
median as $ \ln(\tau^{(255)})$. For large $N$ and small $T$, the
probability distribution of $\tau$ has a very long tail (for large
values of $\tau$), and in many cases we are not able to compute
average values, since for some samples $\tau$ is larger than the
number of sweeps performed.  On the contrary, the median approach
works, and allows a fair estimate. In all cases where we are also able
to estimate the average value of $\ln(\tau)$, we find that it is very similar
to the median value.  Thanks to this approach we have been able to
estimate $\tau_1$ on all our lattice sizes down to $T=0.4$, $\tau_2$
down to $T=0.5$ and $\tau_3$ down to $T=0.6$. Statistical errors have
been computed using the usual bootstrap procedure.

The second decay time of interest is the time scale that governs the
decay of, for example, the square (time-dependent) overlap.  We
monitor the decay of $\overline{<q(0,t)>_J}$ and of

\begin{equation}
\qdc(t)\equiv \overline{<q^2(0,t)>_J-\langle q^2\rangle_J}\ ,
\end{equation}
and we call $\tau_q$ and $\tau_{q_2}$ the time scales that
characterize the short time decay of these objects (see later for
details about the exact definition).

\begin{figure}
  \centering
  \includegraphics[width=0.45\textwidth,angle=0]{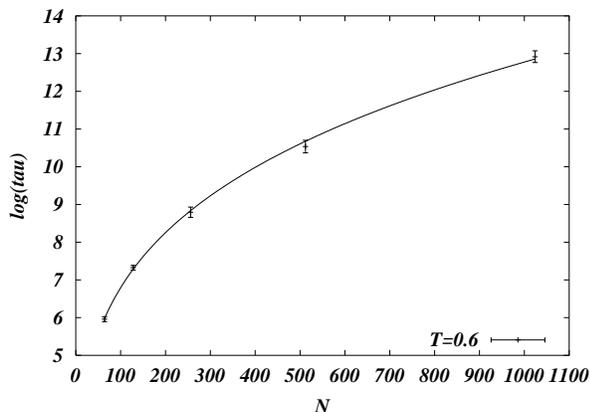}
  \caption[a]{Points with errors are for 
    $\ln(\tau_3)$ versus $N$, and the continuous line for our
    best fit to the form (\ref{E-TAUSCALE}).}  
  \protect\label{F-FLIPTAU}
\end{figure}

Let us start by the results for $\tau_1$, $\tau_2$ and $\tau_3$.  In
figure (\ref{F-FLIPTAU}) we plot one of our most successful fits of
$\tau_3$: here we are at $T=0.6$, the fit is very good and we estimate

\begin{equation}
  \epsilon_{\tau_3}(T=0.6)=0.25\pm 0.04\ , 
\label{E-TAUS}
\end{equation}
that can be compared to

\begin{equation}
  \epsilon_{\tau_1}(T=0.6)=0.20\pm 0.16\ , 
  \epsilon_{\tau_2}(T=0.6)=0.19\pm 0.07\ . 
\end{equation}
Our estimates for $\epsilon_{\tau_1}$, $\epsilon_{\tau_2}$ and
$\epsilon_{\tau_3}$ turn out to be very similar.  The general pattern
that emerges from these fit is of a very good consistency. Let us go
in some more details. Fits to $\epsilon_{\tau_3}$ (here, as we said,
we wait for $q$ becoming negative and equal to $-\Sigma$) are
available only down to $T=0.6$ (at lower $T$ values $\tau_3$ is too
large and we are not able to estimate at all $\epsilon_{\tau_3}$). For
$T$ from $0.6$ up to $0.8$ the best fit is very stable with an
exponent close to $0.25-0.30$. When going too close to the critical
point the behavior becomes less clean.  $\epsilon_{\tau_2}$ (where we
wait for $q$ becoming zero) can be determined down to $T=0.5$ (
$\tau_2$ is smaller than $\tau_3$). Here fluctuations are slightly
larger than in the former case, but again up to $T=0.8$ the exponent
fluctuates in the range $0.2-0.3$.  In the $\epsilon_{\tau_1}$ case
(where we only wait for $q$ decreasing from $1$ down to $+\Sigma$) we
succeed to get a good estimate down to $T=0.4$. Again here, for
example, we estimate $\epsilon_{\tau_1}(T=0.4)\simeq 0.25$, and we get
a quite stable fit in $T$.  We remark that when $T$ approaches $T_c$
the estimates of $\epsilon_{\tau_{1,2,3}}$ have large errors: $\alpha$
becomes very small (one expects $\alpha \to 0$ for $T\to T_c$) and the
leading $N^{\epsilon}$ behavior cannot be distinguished, with the
present range of system sizes, from sub-leading corrections.  It is
also important to notice that our data fully confirm that different
ways to estimate the correlation times (the $1$, $2$ and $3$ $\tau$'s)
lead to the same scaling behavior, with a scaling exponent close to
$\simeq 0.3$.

\begin{figure}
  \centering
  \includegraphics[width=0.45\textwidth,angle=0]{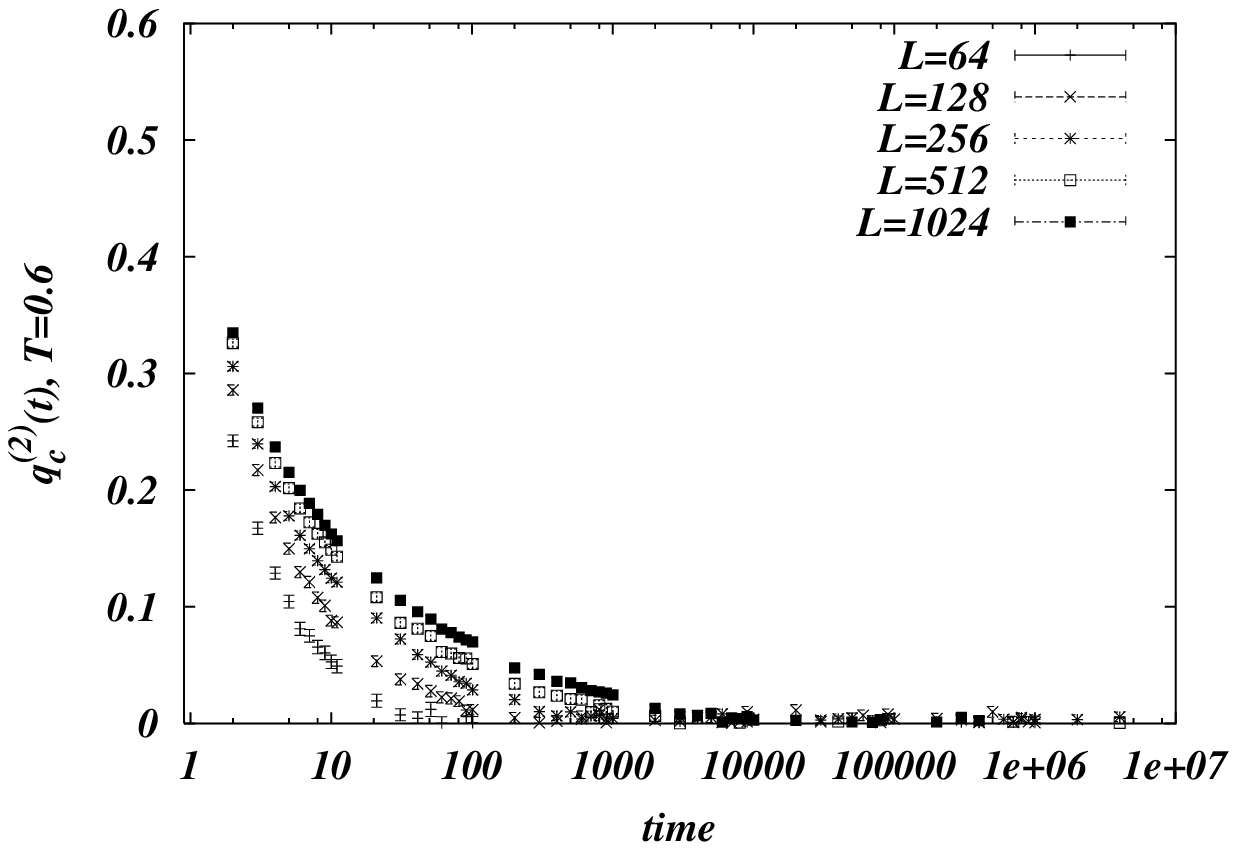}
  \caption[a]{
    $\qdc(t)$ versus $\ln(t)$ at $T=0.6$.
  \protect\label{F-Q2T06}}
\end{figure}

Let us notice here (and this is the focal point of this note, that we
will discuss better in the following) that the result of equation
(\ref{E-TAUS}) {\bf does not manifest}, as opposite to the findings of
\cite{MACYOU}, a scale of the order of
$\exp(c N^{\frac12})$. The scale we observe is governed by an exponent
close to $0.3$.

\begin{figure}
  \centering
  \includegraphics[width=0.45\textwidth,angle=0]{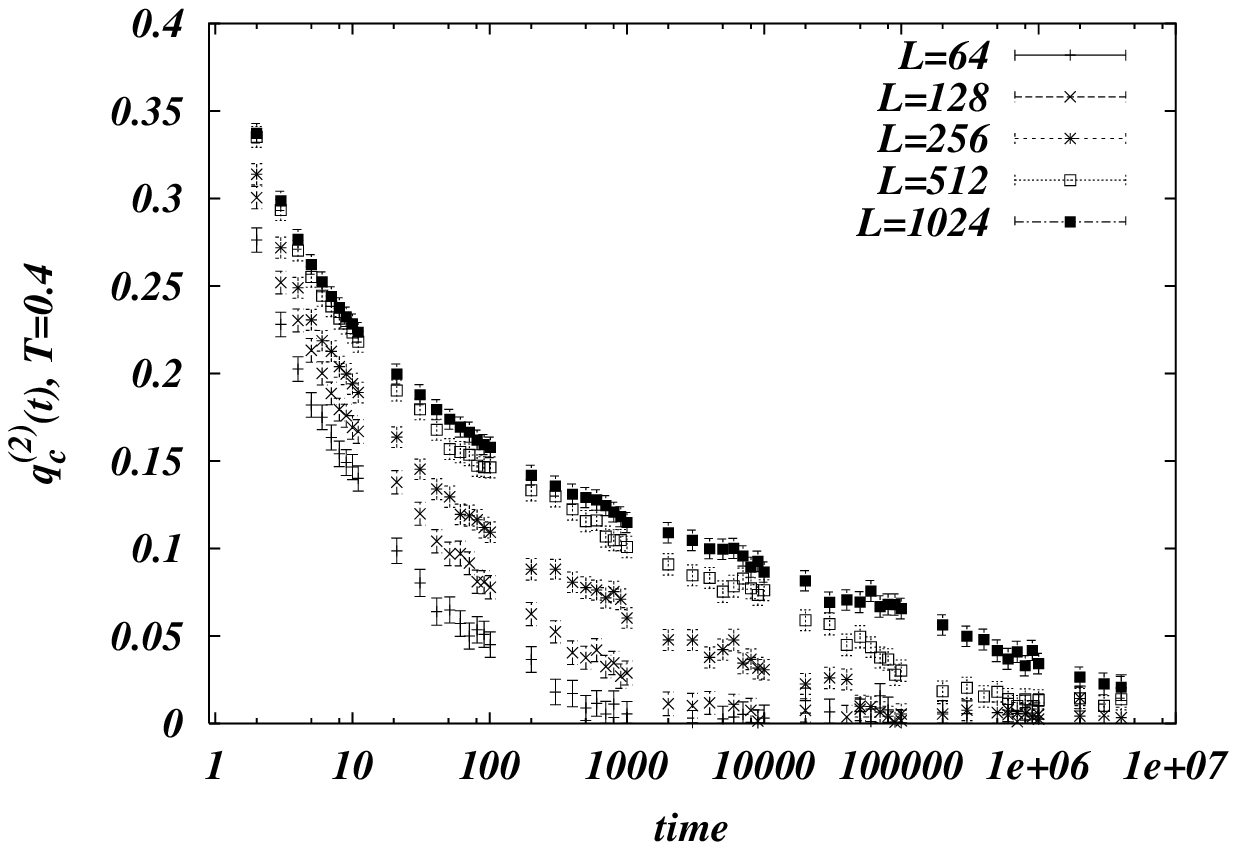}
  \caption[a]{
    $\qdc(t)$ versus $\ln(t)$ at $T=0.4$.
  \protect\label{F-Q2T04}}
\end{figure}

We discuss now the measurements of correlation times that do not
involve the reversal of all the spins. As an example we plot in figure
(\ref{F-Q2T06}) $\qdc(t)$ versus $\ln(t)$ at $T=0.6$, and in figure
(\ref{F-Q2T04}) the same quantity at $T=0.4$.  The two figures exhibit
two regimes separated by some crossover value $t_{max}$: a small time
regime, where $\qdc(t)$ decays slowly with $\ln(t)$, and a large t
regime where $\qdc(t)$ is very small.  This is very suggestive of the
existence of a whole spectrum of relaxation times, up to some maximal
value $\approx t_{max}$.

We have defined the correlation time $\tau_{q_2}$ by computing the
time needed for $\qdc(t)$ to decrease from the value $0.25$ to a
threshold value $\cal T$ that we vary (reference \cite{MACYOU} was
looking directly to the moment in which $\qdc(t)$ is close enough to
zero)\footnote{It is important to note that the ergodic correlation
times $\tau_i$ and these $\tau_{q_2}$, $\tau_q$ are defined in very
different ways, and none of them as a simple, {\it bona fide}
coefficient of an exponential decay $e^{-t/\tau}$.  The fact that we
find that they satisfy reasonable scaling laws shows that the
definitions we use are well founded.}.  In the case of $\qdc(t)$ we
have used the two threshold values ${\cal T}_1=0.125$ and ${\cal
T}_2=0.050$.

\begin{figure}
  \centering \includegraphics[width=0.45\textwidth,angle=0]{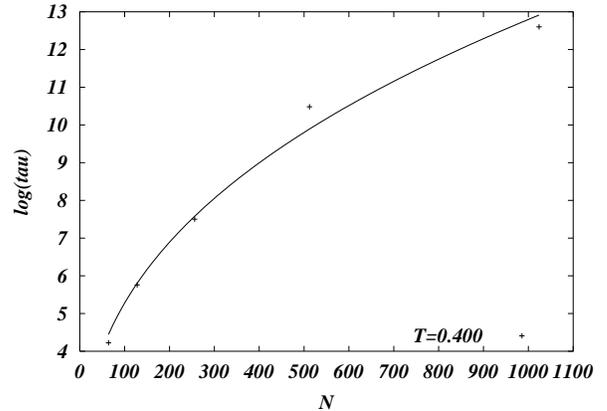}
  \caption[a]{Data points are for $\ln(\tau_{q_2})$ (without
  error-bars) versus $N$, and the continuous line for our best fit to
  the form (\ref{E-TAUSCALE}).  \protect\label{F-Q2TAU}}
\end{figure}

The exponents we estimate by best fits to the form (\ref{E-TAUSCALE})
are again quite stable (even if in this case we have not been able to
produce reliable error estimates) and, let us note right ahead, if any
they are larger than the one estimated for the full reversal times
$\tau_{1,2,3}$: we can  be quite precise on the claim that
the scenario where a slower time scale governs the full spin reversal
while a faster time scale governs the valley to valley migration does
not apply. As an example we plot in figure \ref{F-Q2TAU} the
$\tau_{q_2}$ time as a function of $N$, and our best fit to the form
(\ref{E-TAUSCALE}) at $T=0.4$ and for a threshold ${\cal T}_2=0.050$:
the estimated exponent is here $0.38\pm 0.05$. The exponent values are
very stable when changing the value of the lower threshold, that is a
very good sign. In the $T$ range $0.5-0.8$ the estimated value of
$\epsilon$ are in the range $0.28-0.38$, i.e. completely compatible
with the value $\frac13$ that is reasonable from a theoretical point
of view (see for example \cite{RODMOO,VERVIR}). The quality of the
best fit degenerates again when $T$ becomes too close to $T_c$. It is
maybe worth to stress here that the determination of the exponent
$\epsilon$ is a very difficult problem, exponentially more difficult
than the usual determination of critical exponents, since here instead
of a power behavior we are trying to fit an exponential to a power
behavior: if $\tau$ is ranging over $5$ order of magnitudes (that
would be more than acceptable for a power fit) its logarithm is
ranging over half a decade only, that gives a poor basis for our fit
to the exponential of a power law.

\begin{figure}
  \centering
  \includegraphics[width=0.45\textwidth,angle=0]{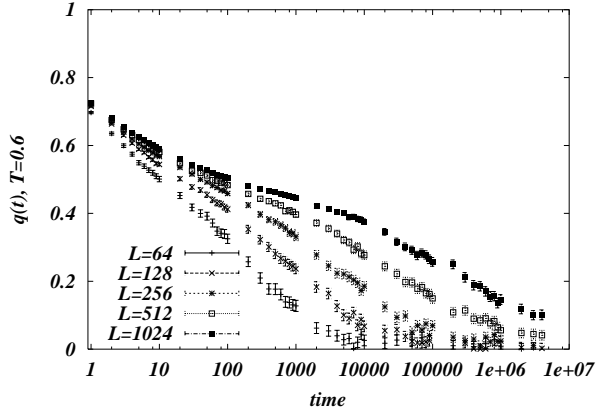}
  \caption[a]{
    $q(t)$ versus $\ln(t)$ at $T=0.6$.
  \protect\label{F-QT06}}
\end{figure}

\begin{figure}
  \centering
  \includegraphics[width=0.45\textwidth,angle=0]{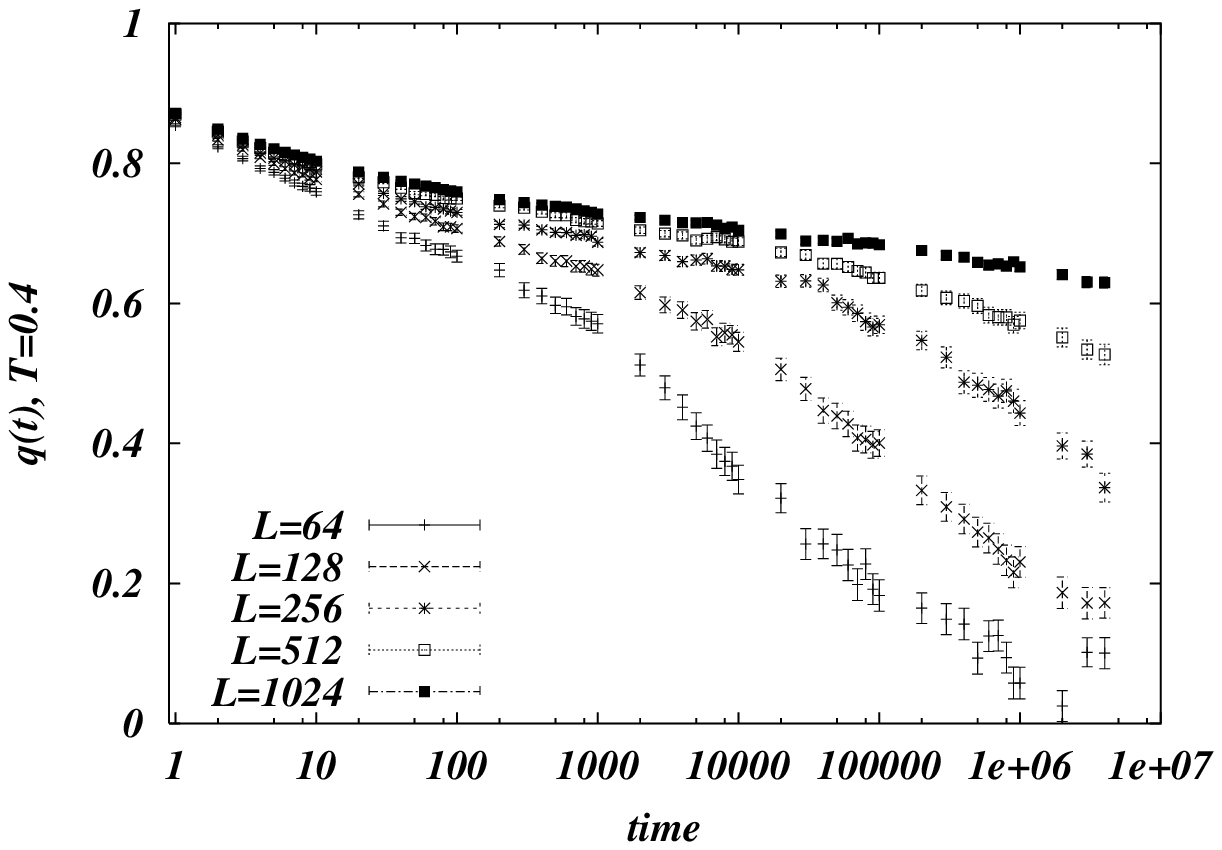}
  \caption[a]{
    $q(t)$ versus $\ln(t)$ at $T=0.4$.
  \protect\label{F-QT04}}
\end{figure}

We have also measured $q(t)$, that we plot in figures \ref{F-QT06} and
\ref{F-QT04} for $T=0.6$ and $T=0.4$ respectively. At large times
$q(t)$ goes to zero,  on the contrary we expect the initial decay
to be governed from the same process that determines the decay of
$\qdc(t)$. It is also interesting to note that we are observing the
expected plateau at the Edwards-Anderson value of the self-overlap,
$q_{EA}$: with good approximation one estimates \cite{MEPAVI}
$q_{EA}(T=0.6)\simeq 0.50$ and $q_{EA}(T=0.4)\simeq 0.74$. These two
values coincide very well with the locations where on our larger
lattice we see a plateau: this is very clear at $T=0.4$ in figure
\ref{F-QT04} and a bit less clean but also evident at $T=0.6$ in
figure \ref{F-QT06}. The finite, large system, spends a long time at
$q_{EA}$ before having $q(t)\to 0$ because of the ergodic transition. 

\begin{figure}
  \centering
  \includegraphics[width=0.45\textwidth,angle=0]{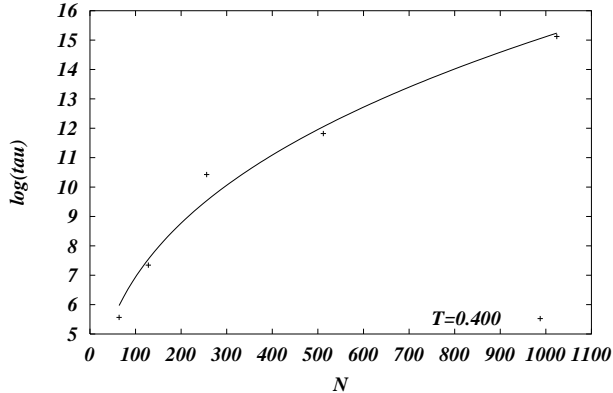}
  \caption[a]{Data points are for 
    $\ln(\tau_{q})$ (without error-bars) versus $N$, and the continuous line for our
    best fit to the form (\ref{E-TAUSCALE}).
  \protect\label{F-QTAU}}
\end{figure}

We have checked that by fitting with the same procedure used for
$\qdc(t)$, using this time the $q$ interval going from $1$ down to
$0.63$ (we use a higher low threshold to stay far from the actual
decay to zero). Things work well, and we fit a scaling exponent for
the correlation times statistically compatible with the one obtained
for $\qdc(t)$. We show in figure \ref{F-QTAU} the analogous of figure 
\ref{F-Q2TAU}, where the best fit gives $\epsilon_{\tau_q}=0.34\pm
0.02$ (again, very well compatible with the value
$\frac13$). Consistent results (slightly lower, of the order of
$0.25$) are obtained at higher $T$ values.

\begin{figure}
  \centering
  \includegraphics[width=0.45\textwidth,angle=0]{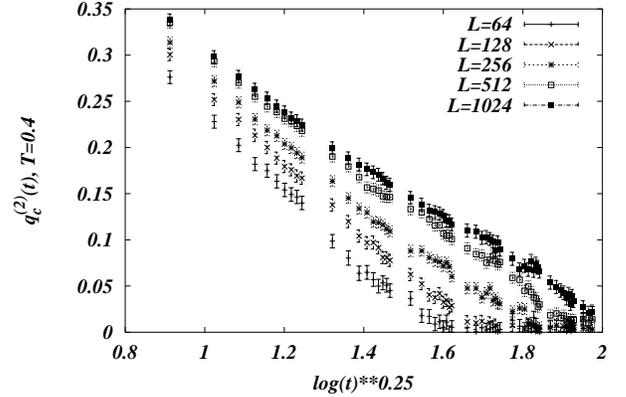}
  \caption[a]{
    $\qdc(t)$ versus $\ln(t)^{0.25}$ at $T=0.4$.
  \protect\label{F-LOGPOWER}}
\end{figure}

It is clear from the figures we have shown that $q(t)$ and $\qdc(t)$
decay very slowly to zero, on a logarithmic scale. We try to be more
quantitative in figure \ref{F-LOGPOWER}, where we show that the
$\qdc(t)$ data are very linear when plotted, for example, as a
function of $\ln(t)^{\beta}$, with $\beta=0.25$: we do not consider
that as a fair determination of $\beta$, since there is a large range
of value of $\beta$ that make the plot linear. What we can claim is
that $\beta$ is surely a small value, of the order of magnitude of
$0.25$. At higher $T$ values we have the same kind of behavior.

\begin{figure}
  \centering
  \includegraphics[width=0.45\textwidth,angle=0]{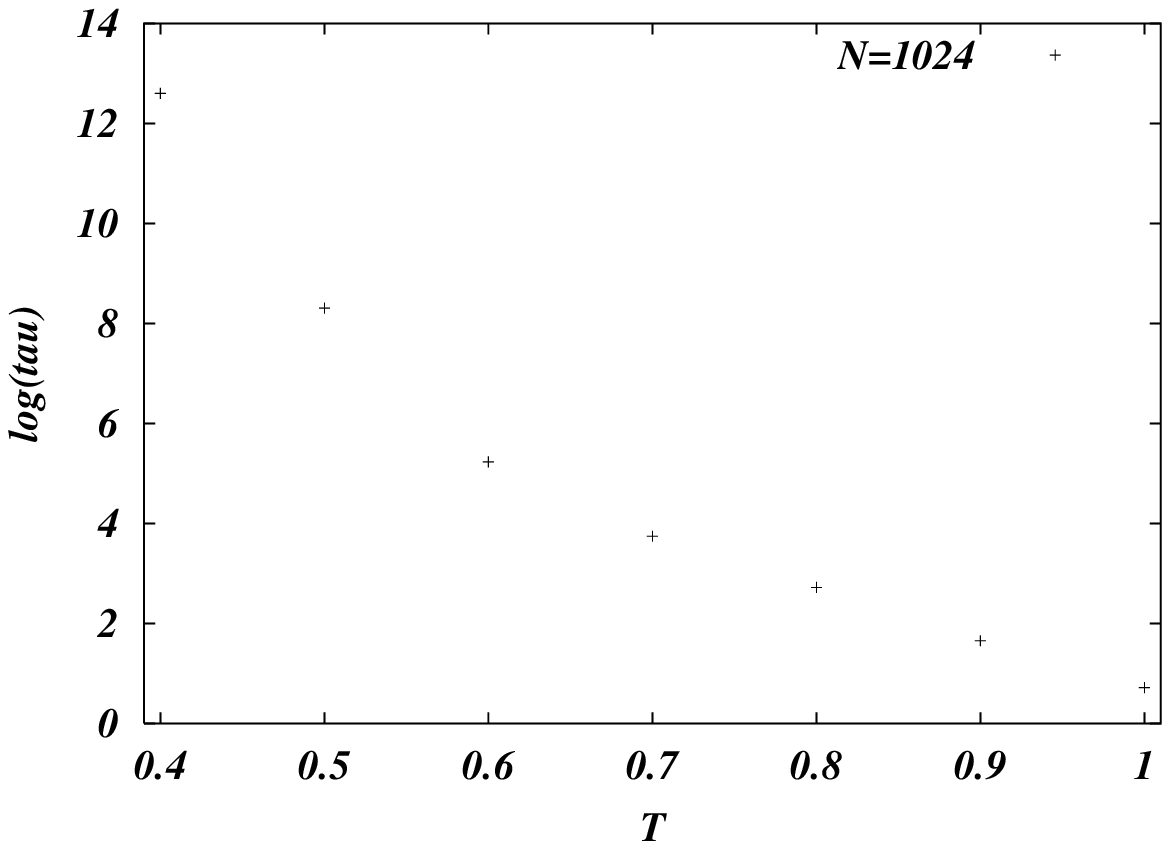}
  \caption[a]{
    $\ln(\tau_{q_2})$ versus $T$ for $N=1024$. The behavior close
    to $T_c=1$ is very linear, while at smaller $T$ values the
    increase of  $\ln(\tau_{q_2})$ becomes sharper.
  \protect\label{F-TAUVST}}
\end{figure}

In figure \ref{F-TAUVST} we show, for our largest system,
$\ln(\tau_{q_2})$ as function of $T$. The data are very well explained
by the fact that we expect an Arrhenius like behavior,
$\exp(\frac{A}{T})$, with $A\simeq (T_c-T)$ \cite{RODMOO}: a
coefficient proportional to $\frac{T_c-T}{T}$ fits indeed the data
very well.

We can sketch a few conclusions.  In the Sherrington-Kirkpatrick mean
field model of spin glasses one single time scaling dictates the
behavior of the correlations times related to the complete reversal of
all spins and to the transitions through the different states that
constitute the phase space: the speculation suggesting that one could
get two different scaling laws is not founded. It is not easy to get
precise values for the exponent that characterizes this exponential
scaling, but all our findings are compatible with a $\epsilon=\frac13$
scaling: this is consistent with barriers scaling like $N^{\frac13}$
\cite{RODMOO,VERVIR}. We have also been able to show that the
connected squared overlap decays to zero with a power of the logarithm
of the order of $0.25$ (and clearly not like a power law).

One of us (E.M.) warmly thanks the {\em Service de Physique
Th\'eorique} of {\em CEA/Saclay} and the {\em Laboratoire de Physique
Th\'eorique et Mod\`eles Statistiques} of {\em Universit\'e Paris-Sud}
for the kind hospitality, during which part of this work was done. We
thank Bernard Derrida, Giorgio Parisi, Felix Ritort and Marta Sales for
useful conversations.

\end{multicols}
\end{document}